\documentclass[runningheads]{llncs}

\usepackage{eccv}

\usepackage{eccvabbrv}

\usepackage{graphicx}
\usepackage{booktabs}
\usepackage{tabularx} 
\usepackage[accsupp]{axessibility}  

\usepackage{hyperref}

\usepackage{orcidlink}
\usepackage{footmisc}

\begin{document}

\title{Facial Expression-Enhanced TTS: \\Combining Face Representation and Emotion Intensity for Adaptive Speech} 

\titlerunning{Facial Expression-Enhanced TTS}

\author{Yunji Chu\textsuperscript{*}\orcidlink{0009-0003-8348-1312} \and
Yunseob Shim\textsuperscript{*}\orcidlink{0009-0002-2565-8100} \and
Unsang Park\textsuperscript{\textdagger}\orcidlink{0000-0001-5539-9520}}

\authorrunning{Y.~Chu et al.}

\institute{Sogang University, Seoul, Korea\\
\email{\{yungie222,yoonseop,unsangpark\}@sogang.ac.kr}}

\maketitle

\begin{abstract}
  We propose FEIM-TTS, an innovative zero-shot text-to-speech (TTS) model that synthesizes emotionally expressive speech, aligned with facial images and modulated by emotion intensity. Leveraging deep learning, FEIM-TTS transcends traditional TTS systems by interpreting facial cues and adjusting to emotional nuances without dependence on labeled datasets. To address sparse audio-visual-emotional data, the model is trained using LRS3, CREMA-D, and MELD datasets, demonstrating its adaptability. FEIM-TTS's unique capability to produce high-quality, speaker-agnostic speech makes it suitable for creating adaptable voices for virtual characters. Moreover, FEIM-TTS significantly enhances accessibility for individuals with visual impairments or those who have trouble seeing. By integrating emotional nuances into TTS, our model enables dynamic and engaging auditory experiences for webcomics, allowing visually impaired users to enjoy these narratives more fully. Comprehensive evaluation evidences its proficiency in modulating emotion and intensity, advancing emotional speech synthesis and accessibility. Samples are available at: \url{https://feim-tts.github.io/}.
\end{abstract}\

\renewcommand{\thefootnote}{\fnsymbol{footnote}}
\footnotetext[1]{\hspace{0.5em}Authors contributed equally to this work.}
\footnotetext[2]{\hspace{0.5em}Corresponding author}

\renewcommand{\thefootnote}{\arabic{footnote}}

\section{Introduction}
\label{sec:intro}

In the rapidly evolving domain of speech synthesis, the enhancement of text-to-speech (TTS) technologies has witnessed significant advancements through the integration of deep generative models. Among these, diffusion models have emerged as a pivotal force, demonstrating unparalleled efficacy in generating high-fidelity images \cite{dhariwal2021diffusion, saharia2022palette}, videos \cite{bar2024lumiere}, and speech \cite{gao2023e3}, thereby revolutionizing acoustic modeling and vocoder technologies \cite{kim2022guided, koizumi2022specgrad}. Concurrently, leveraging facial images to identify speaker-specific attributes in TTS has heralded a paradigm shift towards more personalized and adaptive speech synthesis. Notably, the FACE-TTS model represents a pioneering effort in employing facial cues for zero-shot TTS synthesis, albeit without incorporating emotional nuances in speech \cite{lee2023imaginary}.

Parallel advancements in emotional TTS synthesis have introduced innovative methods for embedding and modulating emotional tones within synthesized speech. This exploration has led to the development of models that facilitate soft-label guidance for fine-grained emotion control, such as EmoDiff \cite{guo2023emodiff}, enabling nuanced manipulation of emotional intensity. Moreover, advancements such as iEmoTTS \cite{zhang2023iemotts} have paved the way for cross-speaker emotion transfer, effectively disentangling prosody and timbre to enhance emotional conveyance. These developments underscore the critical role of emotional expressiveness in augmenting the realism of virtual agents and enriching human-computer interactions.

Despite these significant strides, the synthesis of speech that intricately integrates facial cues with dynamic emotional modulation remains an underexplored avenue. This research introduces the FEIM-TTS framework, an end-to-end TTS model that combines Facial Expression and Emotion Intensity Manipulation to generate speech. This model stands out by modulating emotional intensity and incorporating facial expressions, marking a novel approach to synergizing emotional and facial dimensions in speech synthesis. Distinctively from prior studies in emotional TTS, FEIM-TTS employs Classifier-Free Diffusion Guidance \cite{ho2022classifier} to facilitate both conditional and unconditional training, alongside inference tailored to the intensity of emotions, setting a new precedent in the field.

\section{Related Works}

\subsection{Score-Based Diffusion Model}
Score-based diffusion models \cite{song2020score} are deep learning-based generative models that utilize a gradual process of adding and removing noise to model data distribution.

The forward process is described by a Stochastic Differential Equation (SDE) that gradually adds noise to the data. This noise accumulates in the original data, \(X_0\), creating noisy data \(X_t\), until it reaches complete randomness at \(X_T\). This noise addition process typically uses Gaussian noise and is adjusted according to time or noise level. The forward process is given by:

\begin{align}
  dX_t = \frac{1}{2} \Sigma^{-1} (\mu - X_t) \beta_t \, dt + \sqrt{\beta_t} \, dW_t,
  \label{equation:eq1}
\end{align}
where \( t \in [0, T] \) for some finite time horizon \( T \), \( \beta_t \) is a function that controls the size of the noise, \(\mu\) is a vector, \(\Sigma\) is a diagonal matrix with positive elements, and \( W_t \) denotes a standard Brownian motion.

The reverse process, aiming to predict the original data, can be represented by an SDE as:
\begin{align}
  dX_t = \left( \frac{1}{2} \Sigma^{-1} (\mu - X_t) -\nabla \log p_t(X_t)\right) \beta_t dt + \sqrt{\beta_t} \, dW_t,
  \label{equation:eq2}
\end{align}
where $\Sigma^{-1}$ denotes the inverse of the covariance matrix, $\mu$ represents the mean vector of the data, \( W_t \) represents the reverse-time Brownian motion, and \( p_t \) is the probability density function of the random variable \( X_t \). This SDE is solved backward from the terminal condition \( X_T \). \cite{song2020score} have also shown that an Ordinary Differential Equation (ODE) can be used instead of a \cref{equation:eq2} as follows:
\begin{align}
  dX_t = \frac{1}{2} \left( \Sigma^{-1} (\mu - X_t) -\nabla \log p_t(X_t)\right) \beta_t dt,
  \label{equation:eq3}
\end{align}
Additionally, the neural network \(S(X_t, t)\) estimates the gradient of the log-density of noisy data, \(\nabla \log p_t(X_t)\).

\subsection{Classifier-Free Diffusion Guidance}
Classifier-free diffusion guidance modifies $S(X_t, t)$ to apply classifier guidance without a classifier, by integrating an unconditional and a conditional model within a single framework.

The approach begins with the training of a conditional denoising diffusion model, represented as $p_{\theta}(X | c)$, which is parameterized through a score estimator $S(X_t, t, c)$. In parallel, an unconditional model $p_{\theta}(X)$ is also trained, parameterized by $S(X_t, t)$. The unconditional model uses a null token $\emptyset$ as the class identifier $c$, as shown in the following equation:
\begin{align}
  S(X_t, t) = S(X_t, t, c = \emptyset)
  \label{equation:eq4}
\end{align}

\section{Method}

This section outlines the architecture of the proposed FEIM-TTS model, focusing on its training and inference. Building on the base architecture of FACE-TTS, FEIM-TTS introduces a unique diffusion process designed for emotional speech synthesis. Utilizing principles similar to classifier-free diffusion guidance, it effectively incorporates emotional cues as a conditioning factor in the diffusion process. This allows for nuanced control over speech synthesis, enabling dynamic and context-aware expression of emotions in the generated speech.

\subsection{Conditional TTS Model Training}

In developing FEIM-TTS, our goal was to craft speech rich in emotional nuances. We achieved this through the integration of classifier-free diffusion guidance, drawing from the principles of score-based diffusion models known for their capacity to transform random noise into structured data. This method begins with noise and, through iterative application of learned gradients, progressively refines it into a coherent output. The diffusion process is directed by a score function capable of high-quality sample generation from complex data distributions.

\begin{figure*}
    \hspace*{-0.6em}
    \centering
    \includegraphics[width=1.04\linewidth]{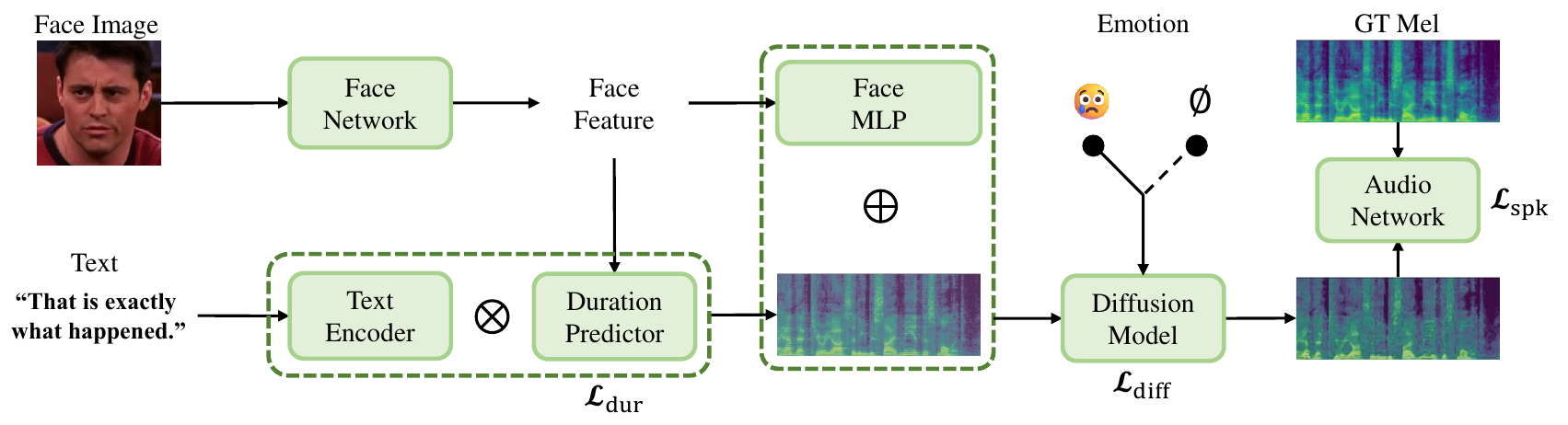}
    \caption{FEIM-TTS Architecture. Leveraging an emotion label, a corresponding facial image, and a textual transcription, our model adeptly synthesizes a mel-spectrogram for expressive speech synthesis. FEIM-TTS, except for the audio network, is trained end-to-end using CREMA-D, MELD, and LRS3 data.}
    \label{fig:Architecture}
\end{figure*}

Classifier-free diffusion guidance enhances our model's generative power by allowing for flexibility in the use of conditioning signals, such as emotion labels, during training. This approach permits the model to generate data both with and without explicit emotional cues, offering control over the emotional tone of synthesized speech.

Emotional embeddings are integrated as conditional inputs, notably with a designated probability of 10\% for applying a null embedding. This strategic probability ensures the model's versatility in generating speech across a spectrum of emotional conditions, from explicitly labeled emotions to more neutral or undetermined emotional states. This capacity for a wide emotional expression range is further supported by training on both labeled and unlabeled data. Our architecture, inspired by FACE-TTS, maintains a focus on leveraging facial images for speaker identity while uniquely incorporating emotional information into the diffusion process. The design and operation of FEIM-TTS are detailed in \cref{fig:Architecture}, highlighting its novel contribution to speech synthesis.

\subsection{Emotion Intensity Controllable Sampling}

During the sampling phase, the model employs a linear combination of conditional and unconditional score estimates, regulated by the emotion intensity parameter. This parameter serves the dual purpose of balancing the quality and diversity of the samples and quantifying the intensity of the emotion being modeled. The prediction of scores occurs in the reverse diffusion process and is formulated as follows, integrating speaker and emotion information into the ODE for reverse diffusion sampling by:
\begin{align}
  dX_t = \frac{1}{2} \left( \mu - X_t - S(X_t, \mu, t, \text{spk}, \text{emo}) \right) \beta_t \, dt
  \label{equation:eq5}
\end{align}
The corresponding equation for the impact of emotion intensity on the linear combination is:
\begin{align}
  w \cdot S(X_t, \mu, t, \text{spk}, \text{emo})-(w - 1) \cdot S(X_t, \mu, t, \text{spk}, \emptyset)
  \label{equation:eq6}
\end{align}

The settings of the emotion intensity parameter are instrumental in modulating the intensity of emotion as:

\begin{itemize}
\item Emotion Intensity at 0: Utilizes null embedding for emotion-agnostic speech generation.
\item Emotion Intensity at 1: Employs input emotion's embedding, focusing solely on that emotion.
\item Emotion Intensity above 1: Amplifies input emotion's effect, producing speech with heightened emotional intensity.
\end{itemize}

This mechanism enables the control of emotion intensity in the sampled speech. Moreover, leveraging speaker information, speech that matches the facial image is generated, ensuring that the output speech is coherent with the visual representation of the speaker. This approach not only enhances the realism and naturalness of the generated speech but also allows for sophisticated control over the emotional expression, offering a flexible framework for creating dynamic and expressive speech content.

\section{Experiments}

\subsection{Experimental Setup}
\begin{table}
  \caption{Overview of emotion distribution, represented by the number of utterances, in CREMA-D and MELD for model training. LRS3, which does not contain emotion labels, is excluded.}
  \label{tab:emotion_distribution}
  \centering
  \begin{tabular}{lccccccc}
    \toprule
    & \textbf{Anger} & \textbf{Disgust} & \textbf{Fear} & \textbf{Happy} & \textbf{Neutral} & \textbf{Sad} & \textbf{Surprise} \\
    \midrule
    \textbf{CREMA-D} & 1,150 & 1,189 & 1,120 & 997 & 940 & 1,159 & - \\
    \textbf{MELD} & 1,128 & 269 & 240 & 1,592 & 4,525 & 720 & 1,099 \\
    \bottomrule
  \end{tabular}
\end{table}

This study leverages three significant English-language datasets to train and evaluate the FEIM-TTS model: CREMA-D \cite{cao2014crema}, MELD \cite{poria2018meld}, and LRS3 \cite{afouras2018lrs3}. The CREMA-D dataset contains 7,442 video clips from 91 actors, delivering 12 distinct sentences across six emotional states—anger, disgust, fear, happiness, neutrality, and sadness—with clips lasting between 2 to 4 seconds. MELD, derived from the TV series ``Friends,'' includes 13,708 utterances across seven emotional categories—anger, disgust, fear, happiness, neutrality, sadness, and surprise—with an average utterance length of 3.59 seconds. These datasets' emotion distribution, crucial for training, is systematically presented in \cref{tab:emotion_distribution}. LRS3, sourced from TED Talks, comprises 31,982 short video segments, each under 100 characters or 6 seconds, devoid of emotional annotations and thus employed for unconditional training aspects of the model.

For the model's conditional training, we partitioned CREMA-D and MELD into 9,989 training, 1,109 validation, and 2,610 testing instances. Conversely, for unconditional training, LRS3 was divided into 31,965 training, 660 validation, and 661 testing instances, highlighting the comprehensive approach adopted for enhancing the model’s ability to generalize across both emotional and natural speech contexts. Notably, the CREMA-D dataset does not encompass the ``surprise'' emotion, a disparity accounted for in our training methodology.

\subsubsection{Data Preprocessing}

To ensure consistency, all datasets were converted to a uniform 16kHz sampling rate. We focused on data segments longer than 2 seconds to improve training efficiency, applying this selection uniformly across all datasets for consistency. Shorter speech segments were excluded only during the diffusion process, while all the utterances were used for other training aspects. Audio features were captured as 128-dimensional mel-spectrograms.

Using the CREMA-D, MELD, and LRS3 datasets, we employed an object detection algorithm \cite{liu2016ssd} to identify faces in video frames. Detected faces were resized to $224\times224$ pixels, and the surrounding area was expanded to enhance facial feature recognition, thereby improving the training phase's effectiveness. Including more of the surrounding context helps the model to learn more discriminative features and reduces false positives by providing a comprehensive context. This approach aligns with findings that deep learning-based face recognition systems benefit from preprocessing steps that retain some surrounding background, which aids in filtering irrelevant information and focusing on relevant facial features \cite{adjabi2020past}.

\subsubsection{Training}

In the initial phase of our model's development, we employed conditional training using the emotionally annotated CREMA-D and MELD datasets. This phase involved a batch size of 64 and spanned 1.43 million iterations. During this period, we encountered noise artifacts in the generated speech outputs, primarily attributed to extraneous sounds and laughter present in the MELD dataset, which is derived from sitcom audio. To address these artifacts, we incorporated classifier-free diffusion guidance and transitioned to an unconditional training phase. For this phase, we utilized the LRS3 dataset, which lacks emotional annotations, as an additional training resource.

The subsequent training on LRS3 was conducted with a batch size of 256 over 22,000 iterations. This strategic combination of conditional and unconditional training significantly improved our model's ability to generate clear and emotionally resonant speech, demonstrating the robustness of our comprehensive training approach within a deep learning framework.

For the model architecture, we selected 128 dimensions for the emotion embedding, 512 dimensions for the face embedding, and used 128-dimensional mel-spectrograms for the audio features. We trained the FEIM-TTS on 8 GPUs (NVIDIA A100 SXM4 80GB) and used the Adam optimizer with a learning rate of 1\(e\)-4.

\subsection{Model Configuration}

The model processes speech, text, and facial images alongside emotions. Facial images are analyzed by the Face Network, and text is processed through the Text Encoder, with both outputs then fed into the Diffusion Model. Emotions guide the diffusion process for noise prediction, also allowing emotion intensity control during inference. The Diffusion Model produces a mel-spectrogram, which, along with the original speech's mel-spectrogram, inputs into the Audio Network.

\subsubsection{Encode}

In Face Network, facial features are extracted from images using a CNN and then combined with text tokens derived from transcriptions. This combination is processed by the text encoder to predict text token durations, assisted by a duration predictor that ensures speech-text alignment. This approach aligns with methodologies proposed by \cite{kim2020glow}.

To enhance the fidelity of this alignment, we focus on optimizing two critical losses: the prior loss and the duration loss. The prior loss refines the model's prediction of text representations by estimating the mean of a normal distribution, thus improving the accuracy of the generated speech's textual content. The duration loss, on the other hand, meticulously adjusts the timing of pronunciation, guaranteeing that the synthesized speech precisely matches the intended text sequences. The integration of these losses is pivotal, significantly boosting the model's effectiveness in producing natural, accurate speech synthesis.

\subsubsection{Decode}

As indicated in \cref{equation:eq1} from Section 2.1, the $\mu$, computed by the text encoder, along with the ground truth (GT) mel-spectrograms, face features, and emotion labels, are provided to the decoder. During the forward diffusion process, the GT mel-spectrogram and $\mu$ are used to generate a noisy mel-spectrogram at time \textit{t}. Subsequently, the noisy mel-spectrogram and facial features are concatenated and input into the noise prediction network. This step is crucial for noise estimation and diffusion loss computation, ensuring accurate and dynamic speech synthesis. Within noise estimation, emotion labels are utilized in the noise prediction network as guiding conditions, similar to how time embeddings are used.

Following the diffusion process, the generated mel-spectrogram and the GT mel-spectrogram are provided to the Audio Network. Within this network, speaker loss, which is a perceptual loss, is calculated by measuring the L1 distance between the features extracted from both inputs.

\begin{figure}
    \centering
    \hspace*{-2.0em}
    \includegraphics[width=1.07\textwidth,height=6cm]{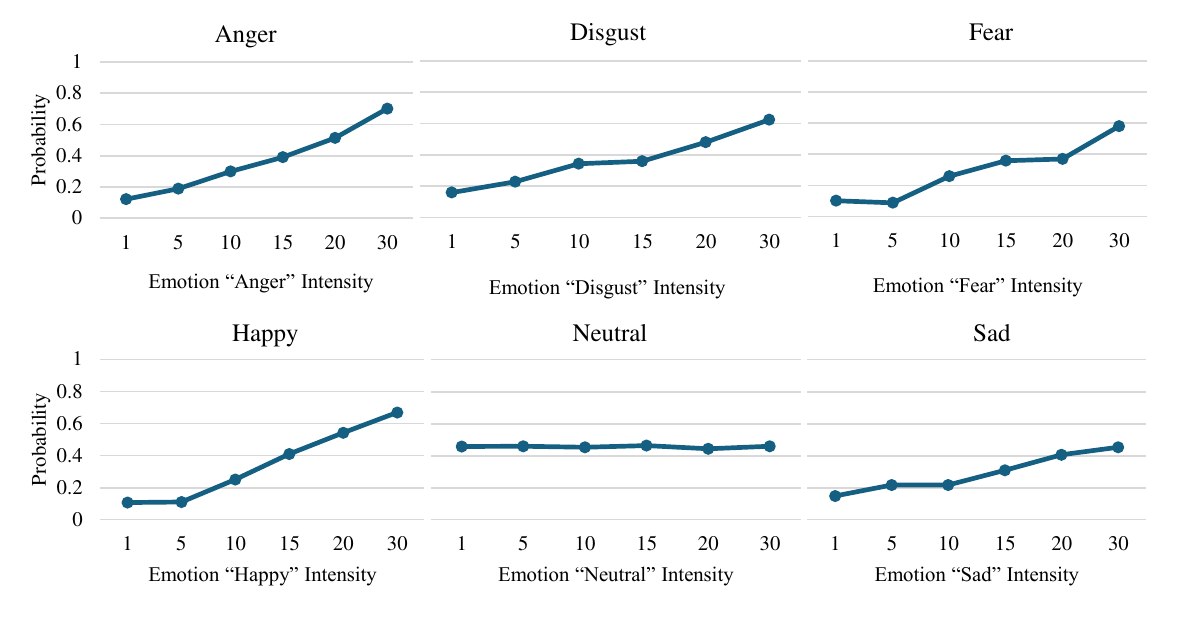}
    \caption{Graph illustrating the relationship between emotion intensity and the class prediction probability of the SER model. This graph shows that lower emotion intensity corresponds to lower prediction probabilities, while higher intensity results in higher probabilities, confirming effective emotion intensity modulation in the synthesized speech of FEIM-TTS.}
    \label{fig:intensity graph}
\end{figure}

\subsection{Evaluations}

\subsubsection{Emotion Intensity Controllability}

To evaluate the capability to control emotion intensity, we employed a Speech Emotion Recognition (SER) framework, specifically the HuBERT model \cite{hsu2021hubert}, fine-tuned on the CREMA-D dataset with a modified prediction layer. This adaptation achieved a test accuracy of 74.98\% on CREMA-D, underscoring its reliability for our evaluation.

We then synthesized 432 speech samples across six distinct emotions and 12 sentences from CREMA-D, employing emotion intensities ranging from 1 to 30. Although theoretically, emotion intensity can be set infinitely high, our empirical observations indicated that pronunciation tends to become slightly muddled at intensities above 30. Therefore, to maintain clarity and quality in our experiments, we limited the range of emotion intensity to 1 through 30.

The fine-tuned SER model's classification accuracy on these samples was 72.22\%, validating the FEIM-TTS model's efficacy in conveying intended emotions.

Further analysis of the class probabilities for each emotion at varying emotion intensities revealed a consistent increase in probability for all emotions, except for Neutral, with higher emotion intensities, as illustrated in \cref{fig:intensity graph}. The Neutral emotion's probability remained largely unchanged, reflecting its characteristic as the baseline emotional state. These findings quantitatively affirm that FEIM-TTS adeptly manipulates emotional intensity in speech synthesis, offering precise control over emotional expression through emotion intensity adjustments.

\subsubsection{TTS Quality and Preference Test}
The evaluation of FEIM-TTS was conducted through a comprehensive set of experiments, incorporating both objective and subjective methodologies. The objective analysis commenced with the computation of Mel Cepstral Distortion (MCD) \cite{kubichek1993mel}, a metric for quantifying the spectral difference between the synthesized speech and ground truth from the CREMA-D dataset. For this assessment, FEIM-TTS was configured with a emotion intensity set to 15, generating 20 speech samples for comparison. The resultant MCD values are detailed in \cref{tab:mcd_values}.

\begin{table}
  \caption{MCD was calculated between the ground truth from CREMA-D and the speech generated by FEIM-TTS, with lower MCD values indicating better performance.}
  \label{tab:mcd_values}
  \centering
  \begin{tabularx}{\textwidth}{X*{5}{>{\centering\arraybackslash}X}}
    \toprule
    & \textbf{Anger} & \textbf{Disgust} & \textbf{Fear} & \textbf{Happy} & \textbf{Sad} \\
    \midrule
    \textbf{MCD↓} & 4.94 $\pm$0.15 & 4.11 $\pm$0.25 & 3.92 $\pm$0.19 & 4.13 $\pm$0.18 & 3.57 $\pm$0.21 \\
    \bottomrule
  \end{tabularx}
\end{table}

\begin{figure}
    \centering
    \includegraphics[width=0.75\linewidth]{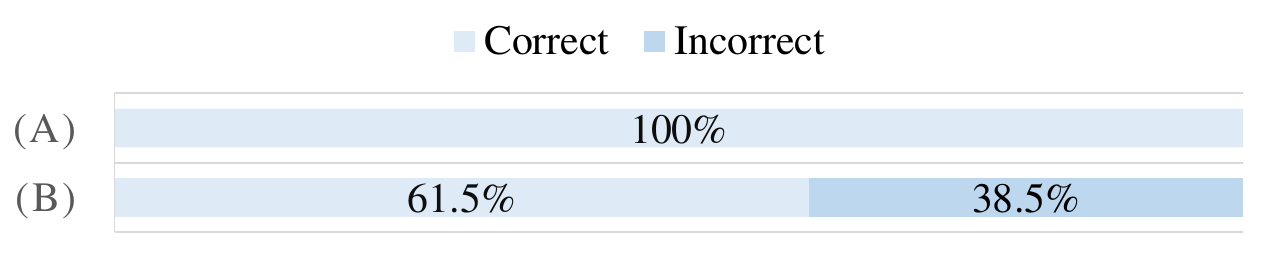}
    \caption{Experimental results were obtained by matching face images with speech. Participants were asked to: (A) Given a face image, generate speech using both the FEIM-TTS and FACE-TTS models, then select which generated speech aligns better with the reference face; and (B) For one of two face images, generate speech using the FEIM-TTS model, then select which of the two face images best matches the generated speech.}
    \label{fig:preference test}
\end{figure}

Subsequent evaluations involved preference tests with 13 participants, designed to ascertain the congruence between generated speech and corresponding virtual facial images. Participants were tasked with identifying the speech sample that most accurately matched a provided virtual face, followed by selecting the most fitting face based on a speech sample. These tests aimed to gauge the alignment between the synthesized speech and visual representation, with findings illustrated in \cref{fig:preference test}.

We also assessed two critical aspects of speech generated by advanced TTS models, FACE-TTS and our proposed model FEIM-TTS. The evaluation was centered around two primary conditions: Intra-Utterance, involving script previously seen during the training of the FEIM-TTS model, and Out-of-Utterance, involving unseen script for both models.

We focused on evaluating the naturalness and suitability of speech generated for virtual face images using two advanced TTS models, FACE-TTS and our proposed FEIM-TTS. The evaluation utilized the Mean Opinion Score (MOS) methodology \cite{streijl2016mean}. As shown in \cref{tab:mos_vs}, in the Intra-Utterance condition, FEIM-TTS significantly outperformed FACE-TTS. In the Out-of-Utterance scenario, FEIM-TTS maintained superior performance for FACE-TTS. These findings underscore the capability of FEIM-TTS to generate more natural-sounding speech, which aligns well with both familiar and novel scripts for virtual faces.

Additionally, a preference test with the same group of participants was conducted to specifically assess how well the synthesized speech matches the virtual faces, further validating the effectiveness of FEIM-TTS over FACE-TTS. This evaluation aimed to determine the degree to which each model's output was deemed appropriate and fitting for the visual representation of the faces. The unanimous preference for FEIM-TTS highlighted its superior capability in aligning speech characteristics effectively with the corresponding facial images, enhancing the overall reality and user experience.

\begin{table}
  \caption{MOS for naturalness in speech generated by two models, FACE-TTS and FEIM-TTS, for virtual face images. Intra-Utterance is a seen script only for FEIM-TTS, while Out-of-Utterance is an unseen script for both models.}
  \label{tab:mos_vs}
  \centering
  \begin{tabularx}{\linewidth}{X*{2}{>{\centering\arraybackslash}X}}
    \toprule
    & \textbf{Intra-Utterance} & \textbf{Out-of-Utterance} \\
    \midrule
    \textbf{FACE-TTS} & 2.62 $\pm$1.15 & 2.15 $\pm$0.66 \\
    \textbf{FEIM-TTS (Ours)} & 4.69 $\pm$0.61 & 3.31 $\pm$1.38 \\
    \bottomrule
  \end{tabularx}
\end{table}

\begin{table}
  \caption{MOS for emotion perception in speech generated by FEIM-TTS for virtual face images and unseen scripts.}
  \label{tab:mos_emotion}
  \centering
  \begin{tabularx}{\linewidth}{X*{6}{>{\centering\arraybackslash}X}}
    \toprule
    \textbf{Anger} & \textbf{Disgust} & \textbf{Fear} & \textbf{Happy} & \textbf{Sad} & \textbf{Surprise} \\
    \midrule
    3.85 $\pm$0.52 & 2.54 $\pm$0.63 & 3.38 $\pm$0.66 & 3.62 $\pm$0.75 & 4.77 $\pm$0.23 & 2.00 $\pm$0.60 \\
    \bottomrule
  \end{tabularx}
\end{table}

Further subjective assessment focused on the emotional expressiveness of the generated speech from the FEIM-TTS model, again utilizing the MOS methodology. Thirteen participants rated the emotional quality of the speech on a scale from 1 (poor quality) to 5 (excellent quality) with an emotion intensity setting of 15 applied to virtual face images and unseen scripts. The MOS outcomes, which included a 95\% confidence interval for six distinct emotions, are compiled in \cref{tab:mos_emotion}. Notably, the synthesized speech achieved an overall average MOS of 3.31. Among the evaluated emotions, ``Sad'' received the highest score of 4.77. However, The absence of ``Surprise'' in the CREMA-D dataset and the limited instances of ``Disgust'' and ``Fear'' in the MELD dataset contributed to their lower MOS scores, highlighting the influence of dataset size and composition on the emotional expressiveness in synthesized speech.

These results comprehensively illustrate the effectiveness of the FEIM-TTS model in generating not only natural-sounding but also emotionally expressive speech, demonstrating its potential for deployment in various applications where dynamic and contextually relevant speech output is essential. The full results and samples of synthesized speech can be accessed at the dedicated project website: \url{https://feim-tts.github.io/}.

\section{Discussion}

In this work, we present FEIM-TTS, a text-to-speech synthesis model that stands out for its sophisticated modulation of emotional intensities within synthesized speech. This model has been rigorously validated through quantitative assessments using an SER model, demonstrating its capability to enhance speech with detailed emotional variations. A hallmark of FEIM-TTS is its ability to accurately link a speaker’s facial features with their speech characteristics, an attribute confirmed through qualitative evaluations involving virtual faces. This feature positions the model as an ideal tool for creating speech for animated characters or virtual avatars.

FEIM-TTS represents a significant stride in the text-to-speech domain by harmonizing facial imagery with emotional content and adjusting emotional intensity. Despite its approach, FEIM-TTS faces challenges in accurately rendering certain emotions, such as ``Surprise'' and ``Disgust,'' due to the limited availability of corresponding data. To address this limitation, we propose augmenting the training process with additional voice-emotion pair datasets like LSSED \cite{fan2021lssed} and RAVDESS \cite{livingstone2018ryerson}, which encompass a broader spectrum of emotional expressions. Our approach includes leveraging an embedding layer within the PyTorch framework to encode emotion labels into embedding dimensions. This method facilitates the integration of novel, previously unlearned emotions by adding their respective embeddings, while retaining and utilizing the pre-trained weights for existing emotions. To further address data imbalance, we generate synthetic speech for underrepresented emotions, thereby enhancing the model's ability to accurately represent a wider range of emotional expressions. This approach is expected to significantly improve the overall performance and generalizability of FEIM-TTS across diverse emotional contexts, making it more robust and versatile for real-world applications.

Moving forward, our research holds significant potential in aiding individuals with visual impairments or those who have trouble seeing. By enhancing Webcomics with dynamic and emotionally rich TTS, our model enables these individuals to enjoy webcomics in a more engaging and joyful manner. The integration of emotional nuances in TTS not only brings characters and narratives to life but also ensures a more immersive experience, thereby bridging the accessibility gap and fostering digital content consumption. This advancement underscores the transformative impact of FEIM-TTS in broadening the horizons of TTS applications, making digital content more accessible and enjoyable for everyone.

\section*{Acknowledgment}

This work was supported by the Institute of Information \& Communications Technology Planning \& Evaluation (IITP) grant funded by the Korea government (MSIT) (RS-2022-II220621, Development of artificial intelligence technology that provides dialog-based multi-modal explainability, 50\%) (RS-2021-II212068, Artificial Intelligence Innovation Hub, 50\%).

\newpage

%
%
\bibliographystyle{splncs04}
\bibliography{main}
\end{document}